# Characteristics of Bi(Pb)–Sr–Ca–Cu–O superconducting ceramics samples properties by method of dynamic magnetic susceptibility measurements

I.A. Yurchenko*, A.V. Nemirovskiy[1)], V.F. Peklun[1)], D.O. Yurchenko

*National Technical University of Ukraine "Kiev Polytechnical Institute",
Faculty of Engineering and Physics,
Department of High-Temperature Materials and Powder Metallurgy,
03056 Kiev, Ukraine*

[1)] *National Technical University of Ukraine "Kiev Polytechnical Institute",
Faculty of Physics and Mathematics,
Department of General and Theoretical Physics,
03056 Kiev, Ukraine*

**Abstract**

Literature and experimental data about practical determination of phase composition and structure of Bi-HTSC ceramics from results of measurements of temperature dependence of magnetic susceptibility are presented. It is shown that Bi-2223 to Bi-2212 phases ratio, calculated by XRD-analysis results, does not correlate precisely with $\chi$–$T$ dependence. Phase, which is present in smaller amount, especially Bi-2212 phase with lower critical temperature ($T_c$), appears less pronouncedly at the $\chi$–$T$ dependence. Besides, limiting parts of Meissner effect manifestation are peculiarities of contacts of superconducting grains. Different kinds of such contacts lead to lowering the $T_c$ together with conservation of narrow transition interval ($\Delta T$), to division of transition into stages, and to increasing of $\Delta T$. Among negative consequences of this influence there are hindrances in realization of grain-oriented Bi-ceramics synthesis scheme which imply powdering of samples consisting of chaotically evolved grains of Bi-2223 phase, ordered placement of obtained plate grains of this phase (for example, through vibration compacting), and sintering.

Keywords*: HTSC; Bi(Pb)–Sr–Ca–Cu–O; synthesis of the 2223-phase; magnetic susceptibility*

## 1. Introduction

Today, high-temperature superconducting materials (HTSC) are represented by ceramic cuprates class. Their structure includes superconducting phase grains with high critical parameters surrounded by weak links system. Grains boundaries contain different defects and impurities. The superconducting grains contacts system can be represented as composite material matrix limiting the superconductor parameters. Besides, HTSC materials are multicomponent superstructures. Mutual substitution of elements, deficit of cations or oxygen, "intergrowth" of phases, defects of structure such as modulations, twins etc., are frequently observed in such materials. Manufacturing conditions variations result in obtaining materials with different characteristics of superconducting grains. Also, grain sizes and their relative orientation (random or ordered) can differ significantly. The combination of mentioned

* *Corresponding author* Permanent address: Department of General and Theoretical Physics, Faculty of Physics and Mathematics, National Technical University of Ukraine "Kiev Polytechnical Institute", 03056 Kiev, Ukraine

material characteristics determines its "quality", i.e. such parameters as superconducting transition temperature and interval, critical current density and its field dependence, critical fields, ability to expel magnetic field (Meissner effect value).

Complex magnetic susceptibility $\chi$ is important characteristic of transition from normal to superconducting state and vice versa. It is a function of temperature and external magnetic field. The observation of superconducting transition of material by magnetic susceptibility measurements allows one to receive more information than by measurements of temperature dependence of electrical resistance. Magnetic susceptibility characterizes all the sample volume. From run of this characterictics, conclusions can be made about structure, appearance of several superconducting and impurity phases, and their arrangement. Besides, magnetic susceptibility measurement does not require rather labour-consuming contacts attachment. Thus, it is nondestructive and rather fast method of material quality control. Below, there is information taken from literature describing the presented measuring method capabilities.

The dependence $\chi$–$T$ is rather sensitive to phase composition of material. If the curve of temperature dependence of real magnetic susceptibility displays several horizontal sections, it can be indicative for presence of several superconducting phases which have different critical temperatures. It is also possible to calculate their relative quantities from this dependence. Quantities of $Nb_3Sn$ (76.6 %) and $Nb$ (23.4%) were determined in [1] from $\chi'$–$T$ dependence for two-phase sample of $Nb_3Sn$. These phases ratio, by electronic scanning microscopy data, is 73.5 % to 26.5%, accordingly. It means that, results of rather fast method of $\chi'$ measurement match those of labour-consuming measurement of the great number of grains sections.

Inflections at $\chi'$ temperature dependence and additional peaks at $\chi''$ one can characterize not only presence of two phases in the material structure but also some other structural characteristics.

Presence of few horizontal levels at $\chi'$–$T$ dependence is most often explained by temperature difference of transitions of superconducting grains and their boundaries. According to the [2], magnetic susceptibility of sintered oxide superconductors consists of two components – intragrain and intergrain. Their relative contributions can be determined changing the modulating field amplitude. The intragrain transition is usually insensitive to small field variation, while the intergrain transition varies even under light field amplitude changes (about $10^{-4}$ mT). The results presented in [3] show that, with the increase of the external constant magnetic field, superconducting transition of sample made from yttrium ceramics is expanded, and the depth of transition of such sample considerably decreases at nitrogen temperatures. Also, temperature of intergrain peak drops down rapidly. Intragrain peak temperature does not change virtually, but rather splits into two peaks. The authors predict that two separate peaks can correspond to grains either with different superconducting parameters, or with different orientation, or to those of different size. In [4], the behavior of bismuth and thallium ceramics in variable magnetic fields with different amplitude was compared. Field amplitude increase results in displacement and extension of transition for both systems. A very small peak is registered at 112 K for bismuth ceramics at maximal field. The authors relate it to 2223 phase grains transition. Apparently, its occurrence is concerned with lower temperatures bias of the basic peak (belonging to grain boundaries). The sample powdering leads to complete disappearance of large peak but does not influence small one. For thallium ceramics the peak appearance under field increase was not observed. However, after powdering of thallium ceramics sample, the much higher peak was observed. The authors think that, at measured dependencies for thallium ceramics, the grain and intergrain peaks superimpose. Rather high peak remained after powdering provides evidence for smaller contribution of weak bonds to superconductivity of thallium ceramics, at least, for samples made by the authors of the abovementioned article.

The authors of [5] distinguish samples with more (with greater $T_c$) or less (with smaller $T_c$) ordered phase by the transition temperature of single-phase samples. For $YBa_2Cu_3O_{7-x}$ films, critical temperature for the first case amounts to approximately 90 K, and for the second one – 86 K. The sample was also manufactured for which three peaks (about 87, 85 and 83 K) were observed. It is known that critical temperature of yttrium ceramics could vary considerably due to oxygen content instability. So, it is obvious for this case that heterogeneity of superconducting film properties resulted in the appearance of transitions at different temperatures.

In [6], an interesting attempt was made to establish the character of two high-temperature superconducting phases distribution in the volume of bismuth ceramics mainly from $\chi'$–$T$ dependence.

From value of imaginary magnetic susceptibility $\chi''$, it is possible to determine loss resistance. There is dependence of maximum $\chi''$ value on grain-oriented yttrium ceramics sample orientation relative to magnetic field is shown in [7]. If sample orientation is parallel to magnetic field, the $\chi''$ value is less than with perpendicular orientation of sample. The authors attribute this to anisotropy of critical currents in the system of weak bonds and different pinning of magnetic vortexes connected with it, which causes power losses. In [8], curves of $\chi$–$T$ dependence of grain-oriented bismuth ceramics are presented for parallel and perpendicular orientation of samples relative to magnetic field lines. The authors consider that, if grains are oriented parallel to the field, they are magnetically "imperceptibly" since their thickness is comparable with London's penetration depth. In this case, only transition of grain boundaries is displayed on the curve of magnetic susceptibility. In case of field direction being perpendicular to grain planes both intergrain and intragrain contributions are presented.

In [9], $\chi$–$T$ dependence curves for composite "yttrium ceramics / polymer" samples with superconducting phase quantity varying in the range of 10–70 % are presented. Grains of superconducting phase in such samples are isolated from each other, which allow authors to conduct series of experiments with purpose of elucidation of intra- and intergrain nature of hightemperature superconductors. Despite zero resistance absence (samples demonstrate typical semiconductor character of $R(T)$ dependence), the diamagnetism is not suppressed completely. The transition depth of composite samples correlates with HTSC-phase content, though, even at 70% phase content, Meissner effect amounts to only about 22% of yttrium ceramics sample transition depth. The authors explain this phenomenon by absence of intergrain bonds contribution to diamagnetism. For composite samples, $T_c^{ons}$ the value remains intact. Increase of AC magnetic field amplitude does not change $\chi'$–$T$ dependence course for composite material, whereas yttrium ceramics sample has shown considerable bias of $\chi'$–$T$ dependence part, related to intergrain weak bonds, to lower temperatures. There is no peak at $\chi''$–$T$ dependence for composite material, whereas for ceramic sample there is rather considerable one. Thus, lower temperatures biasing and broadening of peaks at $\chi''$–$T$ dependence with field increase occur because of vortical losses in superconductor intergrain space. Intragrain diamagnetism, as well as density of critical current, is insensitive to weak magnetic fields. The authors have also determined that sample transition depth ratio at perpendicular and parallel orientation of its grains relatively to magnetic field lines is equal to 1.7–2.0. Concerning influence of superconducting particles size, it is established that its decrease to below 10 μm results in transition depth decrease which is attributed by the authors to depth of magnetic flux penetration into grains. The grain size being increased to above 30 μm, $\chi''$ upgrowth is observed, which is obviously related to vortex motion in coarse grains.

The comparison of superconducting transitions characteristics could be carried out properly only in identical measurements conditions. There are different $\chi'$–$T$ dependencies at different field amplitudes as shown in a number of articles, e.g. [5]. The increase of DC magnetic field amplitude causes the decrease of transition temperature and its "stretching" over the broader temperature range. Besides, this dependence sensibility to transitions of several phases and grains contacts diminishes, i.e., distinguishing of their transition temperatures becomes impossible. As shown in [10], both alternating magnetic field amplitude increase and constant magnetic field growth result in $T_c$ lowering and $\Delta T$ increase. The increase of alternating magnetic field frequency results in $T_c$ increase and $\Delta T$ reduction.

There is much information about critical current determination in HTSC-ceramics through systematic measuring of complex magnetic susceptibility [10–12, etc.]. Physical meaning of $\chi''$ value is loss resistance. The greater $\chi''$ value, on condition that all other parameters values (sample volume, temperature, magnetic field amplitude) are constants, corresponds to the greater area of hysteresis curve, and, therefore, to the greater critical current density. In this case, calculation procedure is based on Been critical state model. According to this model, in the type II superconductors, macroscopic shielding superconducting current density is equal to critical current density. The $j_c$ determination by means of Been method is much easier than transport critical current determination because it is noncontact method. The measurement of high critical current in ceramics by contact method still remains a complex problem.

In conclusion the method for measuring temperature dependence of magnetic susceptibility is quite convenient while investigating technological parameters effect on properties of superconducting ceramics. It allows prompt evaluation of superconducting transition parameters determined by phase composition, structure, and state of grain boundaries. Besides, it is non-destructive method of analysis allowing multiple samples measuring, for example, at different sintering times, or before and after particular processing. However, $\chi'$–$T$ curves interpretation depends on composition and material manufacturing method, therefore additional examinations of particular material specificity are required.

In this work, our own results on measurements of temperature dependence of real magnetic susceptibility of HTSC ceramics Bi(Pb)–Sr–Ca–Cu–O samples are presented, and as well as conclusions concerning in what way various kinds of dependence curves characterize this particular material are made. Our conclusions are based on the comparison of technological features of samples manufacturing and of results of different research techniques (XRD-patterns, electron microscopy, samples mass and sizes change measuring) with the results of $\chi'$–$T$ measurements.

## 2. Experimental

The measuring device ($\nu = 1000$ Hz, $T = 77$–293 K) is based on measurement of Helmholtz coil inductance, which changes under superconducting (i.e., diamagnetic) transition of sample placed inside. The measuring coil is system of two serial flat coils. In the process of estimation of samples quality (combination of phase composition and structure) from this dependence, it is necessary to allow for our device error (about 5 %) in measuring the transition depth along $\chi'$ axis (Meissner effect) at $T < T_{\text{off}}$ which is connected with possible sample displacement in the measuring coils. Therefore we did not pursue the aim of precise quantifying of superconducting phases in sample from $\chi'$–$T$ dependence, however thinking it is possible to estimate their quantities approximately. Within the interval of superconducting transition, temperature measurement error equals to less than 0.3 K which is mainly connected with step-type detection of signal by temperature sensitive unit.

In some cases, $\chi'$–$T$ dependencies, used in [13], are presented as examples describing interrelations of magnetic susceptibility and material properties.

## 3. Results and discussion

*3.1. Determination of phase composition*

Bismuth superconducting ceramics generally consists of two hightemperature superconducting phases which transition temperatures exceed the boiling point of liquid nitrogen (77 K): 2212 – $Bi_2Sr_2Ca_1Cu_2O_{8+x}$ ($T_c \approx 85$ K) and 2223 – $Bi_2Sr_2Ca_2Cu_3O_{10+x}$ ($T_c \approx 110$ K). Besides these phases, ceramics can include simple, double and triple oxides of the basic components. These nonsuperconducting (at $T$ above 77 K) phases can form separate grains, and be distributed as disperse impurities at the superconducting grain boundaries. In this case, the contact of superconducting grains is performed through more or less broad nonsuperconducting interlayer which does, of course, affect the superconducting properties of material. There can be other reasons causing structure aggravation.

When determining phase composition from $\chi'$–$T$ dependence, one can only say about the detection of 2212 and 2223 phases. Nonsuperconducting phases can be determined either as factor causing Meyssner-phase volume decrease, or as presumable reason of transition temperature lowering.

As it is known, hightemperature superconducting phases Bi–2212 and Bi–2223 are "competitive" in a certain sense. Phase 2212 can be formed inside the material under oxide precursor annealing for several hours. When optimal technology is being followed, and the annealing time is increased up to tens or hundreds of hours, phase 2223 is formed from 2212 one. After annealing during more then 500 h, or under other effects influence, phase 2223 disappears, and phase 2212 appears again in ceramics composition. The pattern of these transformations has not been established in details. In a rather broad range of such technological parameters values as components ratio in initial composition, salt precursor decomposition temperature, time and temperature of sintering, etc., phases 2212 and 2223 can coexist in ceramics composition. In the process technology improvement, a possibility to observe two phases appearance in sample, determine bulk content of superconducting phases, and estimate quality of superconducting transition for the both phases is important. Magnetic susceptibility measuring allows to do this. **XRD analysis allows to calculate superconducting phases 2212 and 2223 ratio only. This ratio not necessarily reflects real superconducting properties of material: samples with high 2223 : (2223 + 2212) ratio (above 90%) can demonstrate diamagnetic transition rather far from perfection (extended along temperature axis for tens of К), but diamagnetic transition of samples with 2223 : (2223 + 2212) ratio below 70% can be of rather high quality. Apparently, structural features of ceramics are main factors causing this discordance.**

*3.1.1. Dependence χ′–T for $Bi_2Sr_2Ca_2Cu_3O_{10+x}$ phase sample.* In Fig. 1, *a* temperature dependence of magnetic susceptibility for single-phase 2223-sample is presented. It is evident that superconducting transition of this sample occurs in a narrow temperature range relevant to critical temperature of phase 2223.

It must be noted that *X*-ray patterns showing significant amount of 2212 phase often correspond to such χ′–*T* curves type. However, from the operational point of view, this phase does not show itself as superconductor at working temperature (77 K), i.e. the narrow transition interval can be maintained.

In addition, **for samples with more hightemperature superconducting phase (2223) content predominance over the phase with lower transition temperature (2212) content, it was impossible to observe separate transition for small amount of phase 2212 at *T* > 77 K.** Probably, its transition occurs below the measuring temperatures, since $T_c$ of phase 2212, determined from ρ–*T* dependence (85 K), does not take into account real structure of ceramics and, in case of poor grain contacts (that is quite probable under their small content), can be essentially shifted along *T* axis at χ′–*T* dependence. **Otherwise (at small content of more hightemperature 2223 phase), the transition splits, which was observed for many times.**

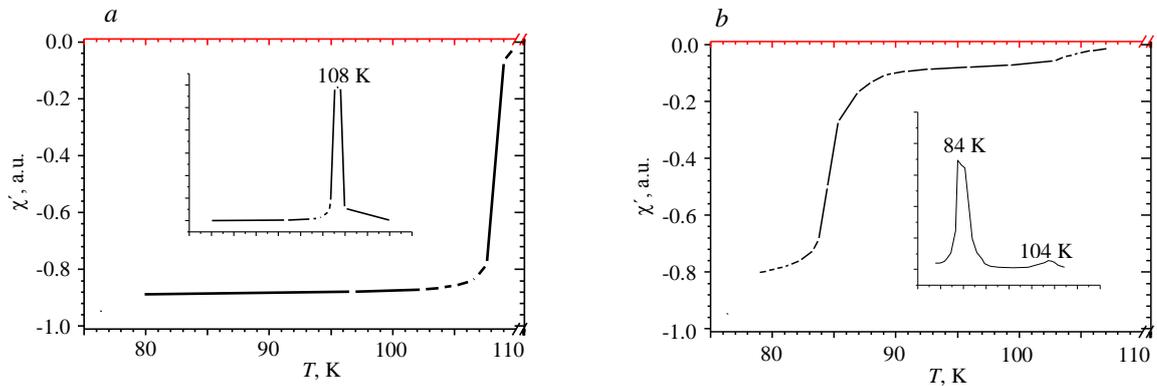

Fig. 1. Temperature dependence of magnetic susceptibility for one-phase (2223) (*a*) and two-phase (2212 + 2223) (*b*) samples

*3.1.2. Dependence χ′–T for two-phase (containing $Bi_2Sr_2Ca_2Cu_3O_{10+x}$ and $Bi_2Sr_2Ca_1Cu_2O_{8+x}$ phases) sample.* From χ′–*T* dependence for two-phase sample (Fig. 1, *b*) the transitions of phases are well parted. The differential curve of χ′–*T* dependence (shown at the insert) allows the transition temperatures to be clearly distinguished. Calculated from the transition depth, HTSC-phases content in this sample makes 7% for phase 2223 and 67% for phase 2212 (or 9.5% and 90.5%, accordingly, with the assumption of 2223 + 2212 = 100%). No conformity of relative phases content calculated from χ′–*T* dependence and from *X*-ray pattern (2223 : (2223 + 2212) = 36%) was observed for the given sample. This also confirms the observation that **the phase presented in smaller quantity less affects the χ′–*T* dependence.**

*3.2. Determination of ceramics structure quality*

Measurements of temperature dependence of magnetic susceptibility allow not only to determine presence of superconducting phases, but also describe material structure. The distribution of nonsuperconducting phases and other structural factors is the relevant parameter influencing the character of superconducting transition at χ′–*T* dependence. The presence of defect areas in superconducting material, especially, probably, in superconduct-ing grain contacts, not only reduces the Meissner effect in proportion to the volume of defects, but also can considerably influence temperature and interval of superconducting transition, down to multiple lowering of the Meissner phase quantity at measuring temperatures exceeding 77 K.

*3.2.1. Dependence $\chi'$–$T$ for the samples in which structural defects do not considerably influence it.* The $\chi'$–$T$ dependencies presented at Fig. 1, *a* and *b*, can serve as the examples of such dependencies for samples with "good" structure.

Superconducting transition of 2223-phase sample with "good" grain contacts takes place at temperatures close to 110 K, and transition interval of such dependence can be 2–3 K (Fig. 1, *a*).

In two-phase sample, quality of dominating phase grains contacts can be rather good, unlike minor phase grains, which cannot be expected to have so good contacts quality. Thus, transition temperature of 2212 phase for the two-phase sample is close to 85 K (Fig. 1, *b*), the interval of this transition being rather narrow. It is impossible to characterize diamagnetic transition of 2223 phase in this sample as high-quality one (because transition interval is insufficiently narrow and $T_c$ is not high).

*3.2.2. Dependence $\chi'$–$T$ for the samples with weak grains contacts.* A concept of the "good" or "qualitative" structure used in this work means that contribution of such components as:
- structural defects of grains of the basic superconducting phases;
- grains contacts of the basic superconducting phases;
- impurities of nonsuperconducting as well as low-temperature superconducting phases;
- porosity;

to structure-sensitive superconducting properties (*SSP*) is minimal.

The presence of porosity and not numerous isolated impurities mainly results in the lowering of superconducting grains volume ratio only, and not in decreasing their contribution to *SSP* (i.e., temperature is not reduced and transition interval is not enlarged). Therefore presence of such defects is not so harmful.

Numerous disperse impurities, which to some extent "isolate" superconducting grains from each other, presence of defects, and absence of structural bonding between superconducting grains in their contact points, especially in case of defect interlayer enlargement or thickening (in case of superconducting grains powdering, and, therefore, of contact contribution increase), reduce superconducting grains contribution to *SSP*.

The presence of superconducting grains structure defects is determined by numerous parameters, e.g. by filling pattern of different elements sublattices, especially, of oxygen sublattice, by "intergrowth" of different homologous *SSP* series members in this system, and by other different structural defects – dislocations, twins, modulations etc. The presence of such defects results in lowering of superconducting grains contribution to *SSP*. The presence of superconducting grains structural defects is substantially determined by technology of superconductor synthesis and can be explored by high-resolution electronic microscopy and by microanalysis. In experimental investigation being the basis for the presented conclusions (which are partly shown in [13]), there is low probability that technological differences could result in considerable differences in superconducting grains structural defects, and, therefore, this problem is not considered in this work (apart from taking into consideration influence of 2223 phase powder additive to an oxide precursor, which was used for synthesis intensification).

Thus, different parameters of superconducting material influence *SSP* as follows:

$$SSP = f\,(bSG;\, A;\, B;\, C),$$

where *SG* – superconducting grains contribution;

*A* – contribution of separate structural defects which do not lower *SG* contribution;

*B* – contribution of structural defects "insulating" *SG*;

*b* – coefficient, connected with *B*, describing the lowering of *SG* contribution owing to their "insulation" ($b < 1$);

*C* – contribution of superconducting grains structural defects.

The *SSP* value, in case of bismuth ceramics, can, by the example of $\chi'$–$T$ dependence, be characterized by integral area under curve of this dependence (Fig. 2) and is determined by degree of approximation of real dependence curve (*2*) by "ideal" curve (*1*).

The presented pattern of role of different structural components is probably somewhat simplified, however, it helps to analyse various experimental curves.

In present work, the interest focuses on the displacement of $\chi'$–$T$ dependence curve to lower temperatures owing to influence of "insulating" structural defects which are characterized by coefficient *b* in the presented scheme.

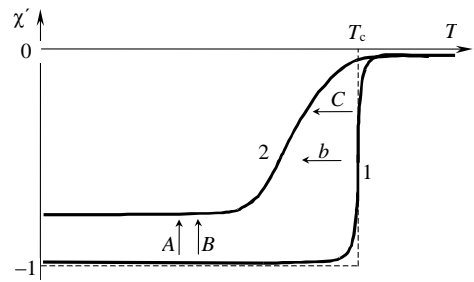

Fig. 2. Some structural parametres influence on magnetic susceptibility vs. temperature trace : 1 – ideal curve; 2 – real curve

The reasons of poor quality of electromagnetic contacts between superconducting grains can be different. It can also affect $\chi'$–$T$ curves in different way. The following varieties of experimental curves are characterized as the result of "insulating" structural defects influence, allowing for samples manufacturing conditions and data of other analytical methods.

*a) $T_c$ lowering at small $\Delta T$ conservation.* Superconducting transition temperature of samples containing only one HTSC-phase can be considerably lowered because of poor material structure.

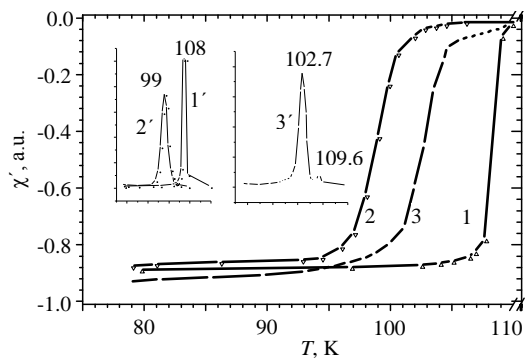

Fig. 3. Real curves illustrated: 1–2 – $T_c$ reduction without significant $\Delta T$ change; 3 – bifurcate transition

Transition temperature of sample No.1 is 108 K, which is close to maximal $T_c$ for this phase (Fig. 3). As it has been previously mentioned, this sample may be considered to have a good structure. Sample No.2 means transition temperature 99 K. Apparently, both transitions correspond to 2223 phase transition. XRD analysis data confirm that 2223 phase predominates in material composition. Transition temperatures, determined from magnetic susceptibility, for the samples made from bismuth ceramics, can range from measuring minimal temperature (77 K) to transition temperature of the most hightemperature superconducting phase 2223 (110 K). Thus, rather legible transitions with a few K $\Delta T$ are observed.

The technological reasons causing $T_c$ lowering at minor $\Delta T$ increase can be specified as follows:

1) addition of 20% and more of powder containing 2223 phase to oxide precursor before sintering [13] (Fig. 4)

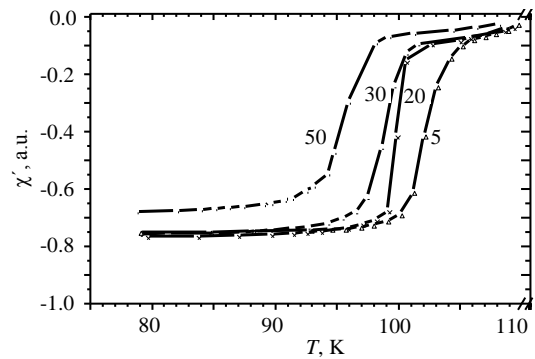

Fig. 4. Reduction of $T_c$ without significant $\Delta T$ change and bifurcate transition for samples with 20% and more 2223-powder additions (some portion of this powder did not react and has poor contacts)

2) grinding of all precursor powder or its part (20% and more) from 30–40 μm down to 3–10 μm [13] (Fig. 5)

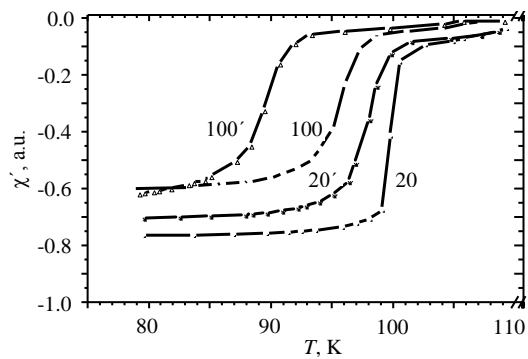

Fig. 5. Reduction of $T_c$ without significant $\Delta T$ change for samples with 20% and more powder additions, grinded from 30–40 to 5–10 μm

3) reduction of sintering time (medium value of $T_c$ for samples with 2223 phase additives (see [13]) after 15 h of sintering is 98 K, after 25 h – 105 K, and after 50 h – 107 K) (Fig. 6). The material structure depends on manufacturing conditions, and it is usually improved with the increase of sintering time. Thus, 2223-phase grains grow and become ordered. The sintering time being increased from 15 to 25 h and more, the "spongy" structure of material is being gradually transformed "plate-like" one.

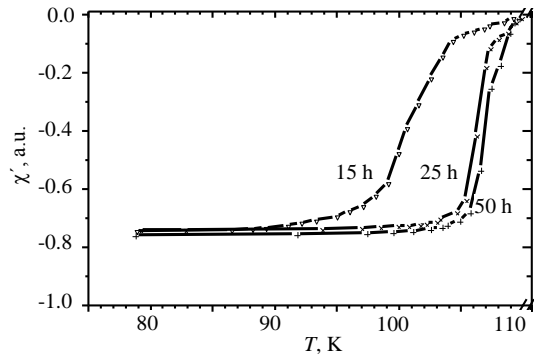

Fig. 6. Reduction of $T_c$ with non-significant $\Delta T$ change for sample sintered during less than optimal (25 h) time

*b) Separation of superconducting transition.* $T_c$ lowering at minor $\Delta T$ increase is supposed to take place only for structurally less qualitative part of the sample. It is possible that, with more precise measurements (for example, at lower field amplitude or at slower change of the sample temperature during measuring), observed curve would be similar to curve No.3 (Fig. 3). According to this curve, the transition of smaller sample volume starts at about 110 K and greater part of the sample transfers to superconducting state at lower temperature, for example, at 102.7 K. Apparently, this sample contains small amount of high-quality superconducting material, but the main part of the sample has "poor" structure. This separation could remain "inconspicuous" in cases, presented in item *a*, for the same reason as transition separation is absent, for example, at phases relation 2223 : 2212 ≈ 70 : 30 (the phase, presented in lower quantity, could not be displayed).

Transition separation in this case is irrelevant of two phases presence, since $T_c$ of 2212 phase (~ 85 K) is much lower than 102.7 K.

The transition separation is often explained by presence of separate transitions for superconducting grains and their contacts. However, the presence of three transition stages is possible [5], which also gives reasons for its separation (there would be more than two structural components). Besides, ratio of superconducting grains and contacts of approximately 10 : 82 (or a little more, if contribution increase of minor phase are taken into account) seems unreal. The mechanical damage of contacts between superconducting grains was observed to suppress manifestation of Meissner effect considerably (item *c*) (see [4, 9] also).

Thus, we have every reason to suppose that, **in case of curves we have obtained, separation of superconducting transition into stages is the result of two structurally distinguished material components presence, each of them including both superconducting grains and their boundaries.** The reason of material inhomogeneity, for example, can be usage of 2223 phase nuclei addition to the oxide precursor (which allows to reduce considerably time of synthesis (see [13])).

*c) Transition stretching to low-temperature area (ΔT increase).* For some samples with large HTSC-phases content, in the range of measuring temperature ($T > 77$ K) no distinct superconducting transition with horizontal level at $T < T_c$ is observed. **In this case there is** probably **gradual superconducting transition, beginning from structurally qualitative areas of material, and ending by less qualitative ones. Thus there is no differentiation of these areas and "stages" on $\chi'$–$T$ dependence are almost not observable.** In such case, determination of superconducting phases quantity from the transition depth is incorrect. The absence of distinct superconducting transition of sample at temperatures above 77 K was observed in the following cases:

1) At examining pelletization pressure effect on properties of bismuth ceramics [14] it was revealed that samples compacted at low pressures increase their weight during sintering approximately by 8 %. At dependence $\chi'-T$ for such samples, transition is extended in a broad temperature range (Fig. 7). Samples compacted at higher pressures have not increased weight during sintering and have shown distinct transition with horizontal level. XRD analysis of all samples shows almost no differences. Presumable reason of poor transition in this case is aggregation of nonsuperconducting phases formed by atoms diffused from filling on grain boundaries (filling powder predominantly contained 2223 phase).

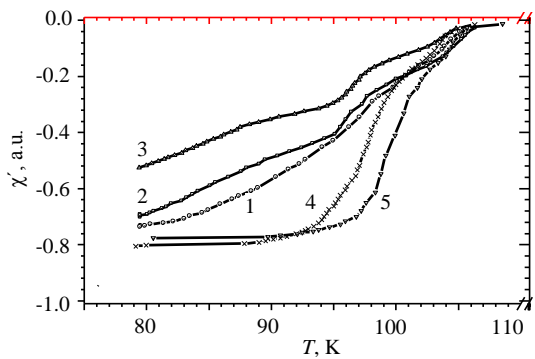

Fig. 7. $\Delta T$ important change in samples compacted at pressure lower than optimal (4–5 t/cm$^2$) and have some quantity of impurities [14]

2) With considerable difference of material composition from obtained phase (2223) composition, arrangement and quantity of impurity phases can be such that superconducting transition does not take place in experimental temperature range ($T > 77$ K). Sample No.1 (Fig. 8, *a*) composition, $Bi_{1,7}Pb_{0,3}Sr_2Ca_2Cu_3O_x$, is close to 2223 phase stoichiometry. It has legible superconducting transition. Sample No.2 composition, $Bi_{1,6}Pb_{0,5}Sr_{2,7}Ca_{2,7}Cu_3O_x$, has some deviation from 2223 phase stoichiometry. The superconducting transition of this sample (at $\chi'-T$ dependence) does not take place in experimental temperature range at all. Curves of $\rho(T)$ dependence are shown in Fig. 8, *b*. From XRD analysis data, 2223 and 2212 phases quantity ratio, 70 : 30, is approximately identical in both samples. These phases predominate in both samples composition. On *X*-ray pattern of sample No.3 one can notice presence of impurity phases considerable peaks, namely, $Ca_2PbO_4$ ($2\theta \approx 17,8°$), and other phases ($2\theta \approx 31,2°$, etc.). Apparently, these phases being formed from excessive components impede the contact of superconducting grain. Probably, quantity of impurity phases in this case is greater than in "low-pressure" samples from previous example. Large quantity of impurity phases transmute them into new quality. The results of electron microscopy demonstrate that structure of this composition samples is much more disperse than that of samples of composition close to 2223 phase. Catastrophic deterioration of dependence $\chi'-T$ was affected by a few factors connected with large quantity of impurity phases influence: unfavorable conditions for HTSC phases growth and their dispersity which, in turn, caused increase of contribution of weak bonds, and also known [9] phenomenon of Meissner effect lowering for grains smaller then 10 μm. In this case it is also possible to suspect the presence of extended superconducting transition at temperatures below liquid nitrogen boiling point as in [15].

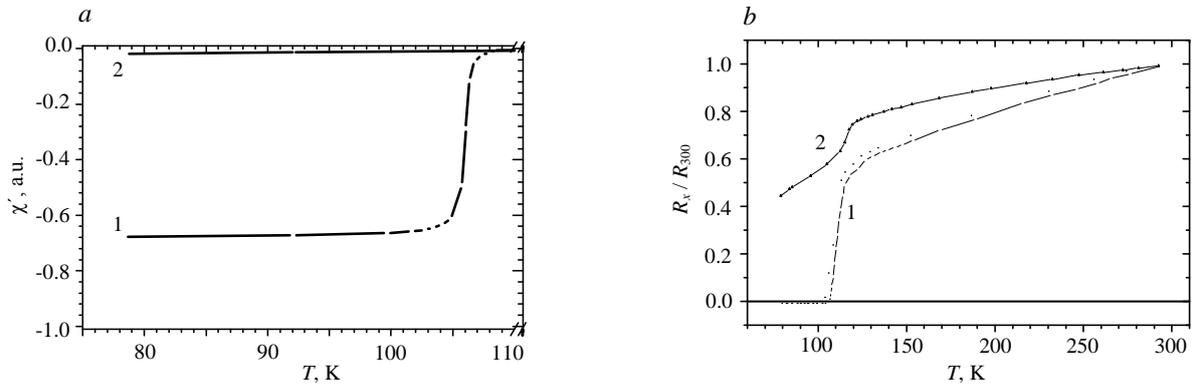

Fig. 8. $\Delta T$ distinguished change from $\chi'$–$T$ (*a*) and only partial transition from $\rho$–$T$ (*b*) of sample (2) which composition is not optimal (it is partially different from Bi–2223 stoichiometry) and contain impurity phases [13]

3) A very weak diamagnetic signal is also observed if 2223-phase sample is grinded (to powder with 30–40 µm particles size) and then compacted again. Such dependence is shown in Fig. 9. Only additional sintering helps to restore grain contacts to some extent. Obviously, after sample grinding, superconducting powder particles are produced, which size, shape, and grain boundaries do not allow obtaining qualitative electromagnetic contact between grains after compaction. In the process of additional annealing, no chemical transformations occur in the material, and contacts are formed between 2223-phase grains, as happens while sintering of other powder systems. However, for full Meissner effect manifestation, several hours of sintering is not enough, even if necks are formed between compacted particles ensuring mechanical contact. Sometimes, even additional sample annealing during 75 h is not enough to eliminate the consequences of superconducting grain contacts destruction completely, though in case of not sufficient grinding of initial sample (to powder with particles size about 30–40 µm) this time is enough for the mentioned purpose.

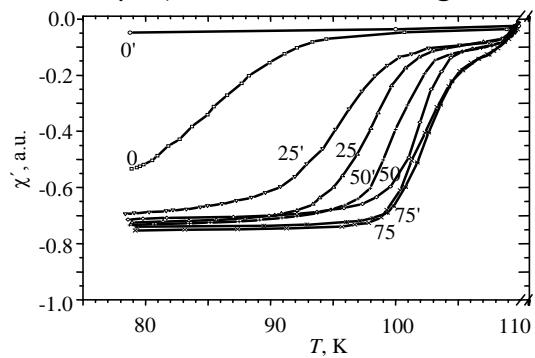

Fig. 9. $\Delta T$ distinguished change of 2223-phase sample which is grinded into powder wiht particle size 30–40 µm and 5–10 µm (numbers with (′)) and then compacted without sintering (0); 25, 50, 75 – time of subsiquent sintering, h

It is worth mentioned that all this conclusions concern only samples of bismuth superconducting system obtained using technology cited in [13]. Features of other HTSC-systems and possible technologies result in fact that such curves are caused by other reasons. So, the authors of [5] can be absolutely right when explaining three stages superconducting transition of $YBa_2Cu_3O_{7-x}$ film by availability of three kinds of superconducting grains. Properties of this superconductor are very sensitive to manufacturing conditions, for example, because of varying of oxygen content. The authors suppose an opportunity of composition fluctuations in the sample area. On the contrary, in case of thallium ceramics, for example, sample destruction and absence of quality contacts between grains do not cause such disastrous degradation of transition as it was shown for the abovementioned example [4]. We have also come across convincing proofs of separation of superconducting transition connected with separation of transition of superconducting grains and their bonds based on dynamics of changes of each χ″ peaks displacement as AC magnetic field amplitude increases.

As a result despite of complexity and some ambiguity of magnetic susceptibility temperature dependence curves explanations, this method is rather demonstrative in technological examinations (when it is important not to obtain an absolute value of characteristics, but rather to compare properties of multiple samples). This dependence gives more complete and complex pattern of superconducting properties, phase composition, and structure interrelation, than temperature dependence of electrical resistance measuring and conventional electron microscopy do. The important advantage of this method is simplicity in realization and does not require contacts attaching and material destruction, and also can be repeatedly applied to the same sample (which is important when studying the dynamics of property changes). The χ′–T dependence measuring may be used as preliminary control method of material structure (relevant for operational characteristics) since this method is more simple than transport critical current measuring or high-resolution microscopy. Besides, through magnetic susceptibility measuring, different physical and material science examinations can be carried out which, however, were not the purposes of the present article (material science aspects are shown in [16]).

## Acknowledgements

We thank N.N. Barkovsky, V.V. Gavriljuk, O.V. Kozlenko, M. Nesterenko for their assistance in measuring temperature dependence of magnetic susceptibility, P. Badica and G. Aldica for measurements of temperature dependence of electric resistance, V.S. Vdovenko for microstructure examination, V.V. Shvajko for realization of XRD.

## References

[1] Y.S. Qian, Y.E. She. Automatic and continuous system for measurement of superconductive transition temperatures using dynamic technique, Cryogenics Vol. 26, No. 4 (1986) 203–206.

[2] H. Mazaki, H. Yasuoka, M. Kakihana, H. Fujimori, M. Yashima, M. Yoshimura. Fundamental and harmonic susceptibilities of an oxide superconductor synthesized by the polymerized complex method, Physica C 252 (1995) 275–281.

[3] G.A. Costa, M. Feretty, G.L. Olcese. On the melt processed $YBa_2Cu_3O_{7-x}$ physicochemical characterization, Solid State Communications Vol. 68, No. 10 (1988) 923–928.

[4] H. Kumakura, K. Togano, K. Takahashi, E. Yanagisava, M. Nakao, H. Maeda. A.C. complex susceptibility measurements in sintered Bi–Sr–Ca–Cu–O and Tl–Ba–Ca–Cu–O superconductors, Japanese Journal of Applied Physics, Vol. 27, No. 11 (1988) L2059–L2062.

[5] E.V. Ekimov, S.I. Krasnosvobodtsev. Investigation of dynamic magnetic susceptibility of superconducting Y–Ba–Cu–O films (in Russian), Superconductivity: physics, chemistry, technique (Moscow), Vol. 6, No. 3 (1993) 577–582.

[6] V.E. Gasumiantc, S.A. Kazymin, V.I. Kaydanov, S.A. Lyikov, V.A. Poliakov, S.E. Khabarov. About an opportunity of definition of features of granular structure bismuthic HTSC-ceramics on their electrical and magnetic properties (in Russian), Superconductivity: physics, chemistry, technique (Moscow), Vol. 4, No. 3 (1991) 586–593.

[7] V.Ya. Malikov, V.P. Seminozhenko, B.L. Timan, R.Yu. Itskovich, L.A. Kvichko, L.A. Kotok, S.E. Logvinova, G.Kh. Rozenberg, P.E. Stadnik. Anisotropy of magnetic properties of textured yttrium HTSC ceramics (in Russian), Superconductivity: physics, chemistry, technique (Moscow), Vol. 6, No. 5 (1993) 1064–1066.

[8] W. Pachla, P. Kovac, H. Marciniak, F. Gomory, I. Husek, I. Pochaba. Structural and electrical properties of Bi(Pb)–Sr–Ca–Cu–O obtained by hot pressing, Physica C 248 (1995) 29–41.

[9] J. Du, J. Unsworth, B.J. Crosby. AC magnetic susceptibility of high-$T_c$ superconducting ceramic / polymer composites, Physica C 245 (1995) 151–158.

[10] S.Y. Ding, J.W. Lin, G.Q. Wang, C. Ren, X.X. Yao, H.T. Peng, Q.Y. Peng, S.H. Zhou. Non-linear vortex dynamics of $(Tl_{0,5}Pb_{0.5})(Ba_{0.2}Sr_{1,8})Ca_2Cu_3O_y$ superconductor in the time window of milliseconds, Physica C 246 (1995) 78–84.

[11] M.W. Lee, M.F. Tai, S.K. Luo, J.B. Shi. Critical current densities in $K_3C_{60}$ / $Rb_3C_{60}$ powders determined from AC / DC susceptibility measurements, Physica C 245 (1995) 6–11.

[12] F. Gomory, P. Lobotka. Determination of shielding current density in bulk cylindrical samples of high-$T_c$ superconductors from AC susceptibility measurements, Solid State Communications, Vol. 66, No. 6 (1988) 645–649.

[13] I.A. Yurchenko, A.F. Alekseev, D.O. Yurchenko, P. Badica, T.Ya.Gridasova, V.V. Morozov, A.V. Nemirovsky, V.F. Peklun. Intensification of Bi–(Pb)–Sr–Ca–Cu–O ceramics synthesis, 384 Physica C (2003) 111–124.

[14] I.A. Fradina, A.F. Alekseev, T.Ja. Gridasova, V.V. Morozov, D.O. Jurchenko. Influence of pelletization pressure on magnetic susceptibility samples of ceramics Bi(Pb)–Sr–Ca–Cu–O, Physica C 311 (1999) 81–85.

[15] P. Badica, K. Togano, H. Kumakura. A modified airtight two-crucible method for growth of Bi-2212 whiskers from glassy pellets. Supercond. Sci. Technol. 17 (2004) 891–898.

[16] I.A. Yurchenko. Representations about nature of processes occurring in the material at different stages of Bi(Pb)–Sr–Ca–Cu–O ceramic manufacture. Journal of Material Science. 41 (2006) 3507